\documentclass[letterpaper,10pt, journal,final]{IEEEtran}      

\usepackage{amssymb}
\usepackage{amsmath}
\usepackage{amsfonts}                                
\usepackage{enumerate}
\usepackage{mathrsfs}
\usepackage{soul,xcolor}
\usepackage{tabulary}
\usepackage{cite}
\usepackage{graphicx}          
\usepackage{color}      
\usepackage{epsfig} 
\usepackage{epstopdf}
\usepackage{times} 
\def\qedp{\hspace*{\fill}~{\tiny $\blacksquare$}}
\setstcolor{red}
\usepackage{soul}
\usepackage{multirow}

\newtheorem{theorem}{Theorem}
\newtheorem{itlemma}{Lemma}
\newtheorem{itdefinition}{Definition}
\newtheorem{itproposition}{Proposition}
\newtheorem{itresult}{Result}
\newtheorem{itremark}{Remark}
\newtheorem{itassumption}{Assumption}
\newtheorem{itcorollary}{Corollary}
\newtheorem{itexample}{Example}

\newenvironment{proposition}{\begin{itproposition}\rm}{\end{itproposition}}
\newenvironment{remark}{\begin{itremark}\rm}{\end{itremark}}
\newenvironment{assumption}{\begin{itassumption}\rm}{\end{itassumption}}
\newenvironment{lemma}{\begin{itlemma}\rm}{\end{itlemma}}

\title{\LARGE \bf  Asymptotic stabilization under homomorphic encryption:\\ A re-encryption free method}

\author{  Shuai Feng, Qian Ma$^*$, Junsoo Kim, Shengyuan Xu
	\thanks{		The material in this paper was not presented in any conference. Corresponding author: Qian Ma.} 
	\thanks{ Shuai Feng, Qian Ma and Shengyuan Xu are with the School of Automation, Nanjing University of Science and Technology, Nanjing 210094, China. Junsoo Kim is with the Department of Electrical and Information Engineering, Seoul National University of Science and Technology, Korea. Emails: 
		{\tt\small s.feng@njust.edu.cn, qma@njust.edu.cn, junsookim@seoultech.ac.kr,  syxu@njust.edu.cn}.}    

}

\begin{document}
	\maketitle    
	
	\begin{abstract}
In this paper, we propose methods to encrypted a pre-given dynamic controller with homomorphic encryption, without re-encrypting the control inputs. We first present a preliminary result showing that the coefficients in a pre-given dynamic controller can be scaled up into integers by the zooming-in factor in dynamic quantization, without utilizing re-encryption. However, a sufficiently small zooming-in factor may not always exist because it requires that the convergence speed of the pre-given closed-loop system should be sufficiently fast. Then, as the main result,
we design a new controller approximating the pre-given dynamic controller, in which the zooming-in factor is decoupled from the convergence rate of the pre-given closed-loop system. 
Therefore, there always exist a (sufficiently small) zooming-in factor of dynamic quantization scaling up all the controller's coefficients to integers, and a finite modulus preventing overflow in cryptosystems. The process is asymptotically stable and the quantizer is not saturated.   
	\end{abstract}

	\section{Introduction}

Industrial control systems are embracing Internet of Things technologies. The low-latency networks, decision-making applications and security technologies will significantly benefit process control in the aspects of promoting efficiency, saving budget and securing data when properly tailored for industrial control systems\cite{sisinni2018industrial,schluter2023brief}. This paper is particularly interested in the security of data in control systems.

Homomorphic encryption (HE) is one of asymmetric encryption schemes that can ensure the confidentiality of data over the communication networks and particularly the controllers. The computation in the controllers can be directly performed over encrypted data without decryption thanks to the homomorphic properties \cite{kim2022comparison, darup2021encrypted}. The authors in \cite{kogiso2015cyber} makes the initial attempt to secure control systems via RSA and ElGamal encryption schemes. Then, various HE-based algorithms have been developed to deal with event-triggered control \cite{shi2024quantization}, model predictive control \cite{darup2017towards}, consensus of multi-agent systems\cite{ruan2019secure}, distributed optimization\cite{alexandru2020cloud}, state estimation\cite{zhang2020secure} and formation control via edge computation\cite{marcantoni2022secure}.

To secure the data of control systems with HE, in general, one needs to carefully tailor the control systems. This is because HE supports the encryption of only integers in a finite set, while the coefficients of controllers, process outputs and control inputs are not necessarily integers. The works \cite{kishida2019encrypted, farokhi2017secure} design encrypted control systems for static feedback controllers via the Paillier cryptosystem. While for the case of designing an encrypted dynamic controller, it is more challenging because one would encounter the overflow issue in general. That is, if the dynamic matrix of a dynamic controller has non-integer elements, one should convert the dynamic matrix to have all integer elements by scaling up. However, the converted controller's state will exceed the finite modulus of a cryptosystem (the size of the message space), and consequently the actuator fails to obtain the correct control inputs by decryption \cite{cheon2018need}. Therefore, the overflow issue in HE-based dynamic control has attracted substantial attention, and the representative solutions such as state reset\cite{murguia2020secure}, designing controllers with integer coefficients\cite{schluter2021stability} and re-shaping the dynamic matrix of a dynamic controller utilizing re-encryption, pole placement and transformation\cite{kimtac1} have been proposed.

In encrypted control, re-encryption refers to a technique in which a third party other than the controller (i.e., actuators) first decrypts some ciphertexts received from the controller, then generates a new state utilizing the decrypted information, encrypts the new state and finally sends the new encrypted state back to the controller \cite{teranishi2023input, kimtac1}. It is clear that a re-encryption process includes decryption and encryption.  
The re-encryption technique mainly requires that the actuators should have both the encryption and decryption capabilities. 
This implies that the overheads of re-encryption might be large if the dimension of the control input is ``large".



In view of the re-encryption technique above, this work is motivated by the following question: given a dynamic controller, how to convert a pre-given controller into a comparable one with only integer coefficients without utilizing re-encryption. The recent papers have proposed re-encryption free methods by reformulating the original controller with non-minimum realization\cite{tavazoei2022nonminimality, lee2023conversion}, and by matrix conversion\cite{kim2021method}. Our paper proposes a new method, which is majorly inspired by \cite{cheon2018need} and \cite{kimcdc}. In \cite{cheon2018need}, the authors have shown that overflow would occur if one scales up the original controller's parameters into integers. This is because the scaling factor also amplifies the controller's state in each iteration step. In our paper, the controller's parameters are also scaled up into integers while overflow problem is prevented. In \cite{kimcdc}, the authors utilize dynamic quantization and modular arithmetic to restore the control inputs from only the lower bits and asymptotically stabilize a process. Meanwhile, the method in \cite{kimcdc} suggests that the zooming-in factor of dynamic quantization should not be too small in average, in order to prevent overflow. Note that this condition is not a problem provided that the dynamic matrix of a dynamic controller has been converted to an integer matrix in advance, e.g., by the results in \cite{kimtac1, tavazoei2022nonminimality, lee2023conversion}.  
In our paper, the zooming-in factor of dynamic quantization is allowed to be sufficiently small even if the dynamic matrix of a dynamic controller contains non-integer elements.

This paper proposes a scheme to encrypt a pre-given dynamic controller with HE, without re-encrypting the control inputs. By the encrypted controller, the control input can be restored by the actuator and the process is stabilized asymptotically. 
We scale up the controller's coefficients into integers by the zooming-in factor in dynamic quantization. First, we present a preliminary result. It shows that if the convergence speed of the original closed-loop system is sufficiently fast, then one can always find a zooming-in factor of dynamic quantization scaling up the controller parameters into integers, and a finite modulus preventing overflow. The preliminary result is consistent with that in \cite{feng2024asymptotic} investigating a tracking control problem by HE, in which the poles of the closed-loop system are designable. However, for a pre-given controller, the preliminary result may not be always applicable because a pre-given closed-loop system may not have a sufficiently fast convergence rate. Then, as the main result, 
we show that by exploiting additional communication resources, there always exists a (sufficiently small) zooming-in factor that scales up a pre-given controller's coefficients into integers by the zooming-in factor. In other words, the zooming-in factor is decoupled from the original control system. The additional cipher messages (from the controller to the actuator) can facilitate the actuator to restore the control inputs asymptotically. This architecture preserves a finite modulus and avoids overflow. Moreover, the process is asymptotically stabilized and the quantizer is not saturated. The explicit lower bounds of the modulus of the cryptosystem and quantization range are provided.   

This paper is organized as follows. In Section II, we introduce the problem formulation, a brief review of re-encryption and control objectives. Section III first presents a preliminary result and then the main result which includes the encrypted control design and lower bounds of the modulus and quantization range. The overheads of re-encryption based and re-encryption free methods are also briefly discussed in Section III.
A numerical example and conclusions are presented in Sections IV and V, respectively.

\textbf{Notation.} We let $\mathbb R$, $\mathbb{Q}$ and $\mathbb Z$ denote the sets of real, rational and integer numbers, respectively. 
For $b\in \mathbb R$, let $\lfloor b \rfloor$ be the floor function such that $\lfloor b \rfloor= \max\{c\in \mathbb{Z}\,|\,c\le b\}$. 
We let $I_n$ denote the identity matrix with dimension $n$ and $\mathbf 0:=[0\,\,0\cdots0]^T$.
For $\chi \in \mathbb{R}$, let $q_{t}(\cdot)$ be the quantization function such that
\begin{align}
q_{t} (\chi)=
\left\{
\begin{array}{cc}
\psi & (2\psi-1)/2 \le \chi  <  (2\psi+1)/2\\
 -q_{t}(\chi)  &  \chi \le - \frac{1}{2}
\end{array}
\right.
\end{align}	
where $\psi =0, 1, \cdots, R$. If $|\chi|<\frac{2R+1}{2}$, we say that the quantizer is not saturated. 
The vector version of the quantization function is defined as $Q(x):= [\,q_t(x_1)\,\,q_t(x_2)  \cdots  q_t(x_n) \, ]^T \in \mathbb Z ^ n$, where $x = [x_1\,\, x_ 2  \cdots x_n]^ T \in \mathbb R ^n$.   
For a vector $x$ and a matrix $\Omega$, let $\|x\|$ and $\|x\|_\infty$ denote the $2 $- and $\infty$-norms of $x$, respectively,  
and $\|\Omega\|$ and $\|\Omega\|_\infty$ represent the corresponding induced norms of matrix $\Omega$. $\rho(\Omega)$ denotes the spectral radius of $\Omega$.

For $g\in \mathbb Z$ and $q\in \{1, 2, \cdots\}$, the modulo operation is defined by $g\,\,\text{mod}\,\,q:=g- \lfloor \frac{g}{q}\rfloor q$.
The set of integers modulo $q\in \mathbb Z_{\ge 1}$ is denoted by $\mathbb Z_q =\{0, 1, \cdots, q-1\}$. For a vector $v=[v_1\,\, v_2  \cdots  v_n] \in  \mathbb Z^n$, $v\,\text{mod}\, q$ denotes the element-wise modulo operation such that $v\,\,\text{mod}\,\, q= [v_1\,\text{mod}\,q\,\,\, v_2\,\text{mod}\,q \cdots  v_n\,\text{mod}\,q]^T \in \mathbb Z _q ^ n$. 
In this paper, by an ``integer matrix" and an ``integer vector", they refer to a matrix and a vector, respectively, whose elements are all integers. An integer vector $v$ can be written as $v= v_1 q + v \,\text{mod}\,q$ in which $v_1$ is some integer vector having the same dimension of $v$ and $q\in \mathbb{Z}$. In our paper, we abuse the notations ``higher bits" and ``lower bits" such that the ``higher bits" refers to $v_1$ and the ``lower bits" refers to $v \,\text{mod}\,q$.

\section{Framework}

\subsection{Preliminaries of additively homomorphic encryption}


In our paper, we focus on additively homomorphic encryption to secure the control systems 


The spaces of plaintexts and ciphertexts are $\mathbb Z_q ^n$ and $\mathcal C^n$, respectively. For a vector in $\mathbb Z_q ^n $, its encryption and decryption processes are given by $\mathbf{Enc}(\cdot): \mathbb Z_q ^n \to \mathcal C^n$ and $\mathbf{Dec(\cdot)}: \mathcal C^n \to \mathbb Z_q ^n$, respectively.
Secret and public keys are also involved in $\mathbf{Enc}(\cdot)$ and $\mathbf{Dec(\cdot)}$, but for simplicity they are omitted. 
The properties of additive homomorphic encryption are listed as follows:\\
\textbf{i.}  For $v \in  \mathbb Z_q ^n$, one has $\mathbf{Dec}(\mathbf{Enc}(v))= v$.\\
\textbf{ii.} Consider the ciphertexts $\mathbf c_1\in \mathcal  C ^n$ and $\mathbf c_2 \in \mathcal  C ^n$.
There exists an operation $``\oplus"  $ such that $\mathbf{Dec}(\mathbf c_1 \oplus \mathbf c_2)= \mathbf{Dec}(\mathbf c_1)+ \mathbf{Dec}(\mathbf c_2) \,\, \text{mod}\,\,q$. \\
\textbf{iii.} Consider a matrix in plaintext $M \in \mathbb Z ^{m \times n}$ and a vector in ciphertext $\mathbf c_3 \in  \mathcal C ^n$. There exists an operation ``$\cdot$" representing multiplication such that $\mathbf{Dec}(M \cdot \mathbf c_3)= M \mathbf{Dec}(\mathbf c_3) \,\, \text{mod}\,\,q$.  

Property ii is the key feature of additively homomorphic encryption schemes such that operating the ciphertexts $\mathbf c_1$ and $\mathbf c_2$ by ``$\oplus$" will lead to the addition of their plaintexts, under the modulo of $q$. 
For more information about additively homomorphic encryption algorithms and their properties, we refer the readers to the survey papers \cite{kim2022comparison, schluter2023brief, darup2021encrypted} and \cite{Paillier}.

\subsection{Problem formulation}

We consider a process described by a discrete-time system
\begin{subequations}\label{process}
\begin{align}
	x_p(t+1)&=A x_p(t)+B u(t)\\
	y_p(t)&=C x_p(t)
\end{align}
\end{subequations}
where $x_p(k)\in \mathbb{R}^{n}$ denotes the state of the process, $u(k) \in \mathbb{R}^{w}$ denotes the control input and $y_p(k)\in \mathbb{R}^{v}$ denotes the output. We assume $A \in \mathbb Q^{n \times n}$, $B \in \mathbb Q^{n \times w}$ and $C\in \mathbb Q^{v\times n}$.
We also assume that the initial condition $x_p(0)$ is not infinitely large and one can find a real number $C_{x_p(0)}$ such that $\| x_p(0)\|_\infty \le C_{x_p(0)}$. 

In this paper, we consider a general class of controllers in the form
\begin{subequations}\label{first controller}
\begin{align}
	x(t+1)=Fx(t)+Gy_p(t) + R r(t)\\
	u(t)=Hx(t)+Jy_p(t) + S r(t)
\end{align}
\end{subequations}
in which 
$x(t)\in \mathbb{R}^{n_x}$ denotes internal state of the controller, $r(t)\in \mathbb{R}^{n_r}$ denotes the reference provided by a third party. 

\begin{assumption}\label{ass 1}
We assume that $F\in \mathbb{Q}^{n_x \times n_x }$, $G\in \mathbb{Q}^{n_x\times v}$, $R \in \mathbb{Q}^{n_x\times n_r}$, $H\in \mathbb{Q}^{n_ w\times n_x  }$, $J\in \mathbb{Q}^{w \times v}$, $S \in \mathbb{Q}^{w \times n_r}$ and $x(0)\in \mathbb{Q}^{n_x }$ in (\ref{first controller}) are pre-given, and the spectral radius of the closed-loop system satisfies
\begin{align}\label{rho_c}
	\rho\left(\begin{bmatrix}
		A+BJC & BH\\GC & F
	\end{bmatrix}\right)=:\rho_c <1.  
\end{align}
\end{assumption}

\textbf{A brief review of the re-encryption of the control inputs \cite{kimtac1}:} 
In \cite{cheon2018need}, the authors find that if $F$ has non-integer elements, the controller is subject to overflow problems. 
To solve the problem, \cite{kimtac1} exploits pole placement, re-encryption and matrix transformation to obtain an integer dynamic matrix as the new dynamic matrix.  
For simplicity, we assume that all the entries in $G, R, J$ and $S$ in (\ref{first controller}) are 0. Assuming $(F, H)$ observable, (\ref{first controller}) can be written as
\begin{subequations}\label{example}
\begin{align}
	x(t+1)&=(F-L_cH) x(t) + L_c u(t)\\
	u(t)&=Hx(t).
\end{align}
\end{subequations}
in which $L_c$ is chosen such that $F-L_cH$ has only ``integer eigenvalues", i.e., $a_i \pm b_i i$ and $a_i \in \mathbb Z$ and $b_i\in \mathbb Z$. 
By letting $x_c(t) := T_c ^ {-1} x(t) / s_c$ and $u_c (t) := u(t)/(s_c s_u)$, (\ref{example}) can be transformed into
\begin{subequations}\label{re}
	\begin{align}
	 	x_c(t+1)&= T_c ^{-1}(F-L_cH)T_c   x_c(t) + \frac{T_c ^{-1} L_c}{s_c} Q(s_cs_u u_c(t))\\
		u_c(t)&=\frac{HT_c}{s_u}x_c(t).
	\end{align}
\end{subequations}
In (\ref{re}), $ T_c ^{-1}(F-L_cH)T_c$ is the Jordan form of $F-L_cH$, which is an integer matrix because $F-L_cH$ has only integer eigenvalues. Moreover, $ T_c ^{-1} L_c/s_c$ and $ HT_c/s_u$ are integer matrices under small $s_c$ and $s_u$, respectively. Then, the controller in (\ref{re}) can be encrypted by HE. However, the ciphertexts of $ Q(s_cs_u u_c(t))$ cannot be generated by the controller. To solve the problem, re-encryption is proposed such that the actuator first decrypts the ciphertext of $u_c(t)$ (received from the controller), encrypts $Q(s_cs_u u_c(t))$, and sends the ciphertexts of $Q(s_cs_u u_c(t))$ to the controller. 

\textbf{Control objectives:} The paper has three control objectives:
\begin{itemize}
\item[1)] For a pre-given dynamic controller, design an encrypted controller over homomorphically encrypted data without re-encryption of the control input. 
\item[2)] The actuator can restore the control input $u(t)$ in (\ref{first controller}) asymptotically under a finite modulus $q$ (the size of the message space of a cryptosystem). 
\item[3)] The process is asymptotically stable and the quantizer is not saturated.  
\end{itemize}  

\section{Controller design} 
In this section, we first present a preliminary result and then the main result of the paper. 

\subsection{A preliminary result}

To operate (\ref{first controller}) over homomorphically encrypted data, quantization is necessary because the plaintexts to be encrypted must be integers. 
Then, the dynamics of the controller (\ref{first controller}) can be approximated by
\begin{align}\label{5}
	\!\! \left\{
	\begin{array}{l}
		x(t\!+\!1)=Fx(t) \!+\! l(t)GQ\left(  \frac{y(t)}{l(t)} \right)  \!+\! l(t)R Q\left( \frac{r(t)}{l(t)}\right) \\
		u(t)=Hx(t)+l(t) JQ\left(  \frac{y(t)}{l(t)} \right)   +l(t) S Q\left( \frac{r(t)}{l(t)}\right)  
	\end{array}
	\right.
\end{align}
in which $l(t)$ updates as
\begin{align}\label{scaling}
	l(t+1)= \omega l(t), \,\,  0<\omega<1. 
\end{align}
In this subsection, for simplicity, we assume that the quantizer $Q(\cdot)$ in (\ref{5}) has infinite capacity and thus is free from saturation. This assumption will be relaxed in Section III-B. 
In (\ref{scaling}), $\omega$ is the so-called zooming-in factor in dynamic quantization, with which one can stabilize the process asymptotically. 



Defining $	\bar x_p(t):=\frac{x_p(t)}{l(t)}, \bar y_p(t):=\frac{y_p(t)}{l(t)},
		 \bar r(t):= \frac{r(t)}{l(t)}, 
	\bar  x(t):=\frac{x(t)}{l(t)}$ and $ \bar u(t):=\frac{u(t)}{l(t)} 
$,
 the dynamics of the process (\ref{process}) is transformed into
$
		\bar x_p(t+1)=\frac{A}{\omega} \bar x_p(t)+\frac{B}{\omega}  \bar u(t) $ and $
		\bar y_p(t)=C \bar x_p(t).
$
In particular, the dynamics of the controller (\ref{5}) is transformed into
$		\bar x(t+1)=\frac{F}{\omega} \bar x(t)+\frac{G}{\omega}Q(\bar y_p(t)) +  \frac{R}{\omega} Q\left(\bar r (t) \right) $ and $  
		\bar u(t)=H \bar x(t)+JQ(\bar y_p(t))+ S Q\left(\bar r (t) \right)$, in which $H, J$ and $S$ are not necessarily integer matrices.
Thus, through defining
$
	\tilde x(t)  : =\frac{\bar x(t)}{s_1}$ and $\tilde u(t)  =\frac{\bar u(t)}{s_1 s_2}, 
$
one can further obtain:
\begin{subequations}\label{fourth controller}
\begin{align}
	\tilde x(t+1)&=\frac{F}{\omega} \tilde x(t)+\frac{G}{s_1\omega}Q(\bar y_p(t))  +  \frac{R}{s_1\omega} Q\left(\bar r (t) \right)   \\
	\tilde u(t)&=\frac{H}{s_2}  \tilde x(t)+\frac{J}{s_1s_2}Q(\bar y_p(t))+\frac{S}{s_1s_2}Q(\bar r(t)). 
\end{align}
\end{subequations}

Let $s_F$ denote the largest positive real number such that $F/s_F$ is an integer matrix: 
\begin{align}\label{definition}
s_F:=\max\{ a \in (0,1]\cap\mathbb Q\,| \, F/a \in \mathbb Z ^{n_x \times n_x}\}.
\end{align}
The following result presents the conditions of $\omega$, $s_1$ and $s_2$ such that (\ref{fourth controller}) contains only integer matrices.  

\begin{lemma}\label{lemma}
Consider $\rho_c$ in (\ref{rho_c}) and $s_F$ in (\ref{definition}). If $\rho_c <  s_F$, then (\ref{fourth controller}) contains only integer matrices by the following steps:\\
(a) select $\omega \in (\rho_c, s_F] \cap \mathbb Q$ such that $\frac{F}{\omega} \in \mathbb Z ^{n_x \times n_x} $, \\ 
(b) select a sufficiently small $s_2>0$ such that  $\frac{H}{s_2} \in \mathbb Z ^{n_w \times n_x} $,\\  
(c) select a sufficiently small $s_1>0$ such that $\frac{G}{s_1\omega} \in \mathbb Z ^{n_x \times v} $, $\frac{R}{s_1\omega} \in \mathbb Z ^{n_x \times n_r}$, $\frac{J}{s_1s_2} \in \mathbb Z ^{w \times v} $ and $\frac{S}{s_1s_2} \in \mathbb Z ^{w \times n_r} $.  \qedp  
\end{lemma}

The proof is straightforward and omitted. In Lemma \ref{lemma}, the condition $\rho_c < s_F$ is very important for ensuring a finite $q$, which will be explained later in Remark \ref{remark 1}. For (a) in Lemma \ref{lemma}, one can find at least an $\omega \in (\rho_c, s_F] \cap \mathbb Q$ because one can always let $\omega=s_F$ and hence $F/\omega$ is an integer matrix by the definition in (\ref{definition}).

Following Lemma \ref{lemma}, (\ref{fourth controller}) contains only integer matrices. Moreover, under $\tilde x(0)= \frac{x(0)}{s_1l(0)} \in \mathbb Z ^{n_x}$ by choosing a small $l(0)$, one can infer that $\tilde x(t)$ and $\tilde u(t)$ are ``integer vectors" for all $t$.  
Then the dynamics of (\ref{fourth controller}) over $\mathbb Z_q$ can be described by 
\begin{subequations}\label{fifth controller}
	\begin{align}
	&\!\!\!	\tilde x^q(t\!+\!1)\!=\!\frac{F}{\omega}\tilde x^q(t)\!+\!\frac{G}{s_1\omega}Q(\bar y(t))  \!+\!  \frac{R}{s_1\omega} Q(\bar r (t) )\, \, \text{mod}\,\, q   \\
		&\tilde u^q (t)\!=\!\frac{H}{s_2}  \tilde x^q (t)\!+\!\frac{J}{s_1s_2}Q(\bar y(t))\!+\!\frac{S}{s_1s_2}Q(\bar y(t))\,\, \text{mod}\,\, q  
	\end{align}
\end{subequations}
with $ \tilde x^q (0)   = \tilde x(0)\, \text{mod}\, q$.
One can encrypt (\ref{fifth controller}) by the HE scheme in Section II-A: 
\begin{subequations}\label{5.5 controller}
\begin{align}
	\mathbf {   x  }(t+1)&= \mathbf {   F \cdot  x  }(t) \oplus \mathbf {   G  \cdot y  }(t)   \oplus   \mathbf {   R \cdot  r  }(t)\\
	\mathbf { u  }(t)&= \mathbf {   H  \cdot  x  }(t) \oplus \mathbf {   J \cdot  y  }(t) \oplus   \mathbf {   S  \cdot  r  }(t)
\end{align}
\end{subequations}
with $\mathbf {  x(0) } = \textbf{Enc}(\tilde x^q(0))$ and 
$	\mathbf {  y(t) } = \textbf{Enc}(Q(\bar y(t)) \,\text{mod}\,q)$, $ \mathbf {  r(t) } = \textbf{Enc}(Q(\bar r(t)) \,\text{mod}\,q)$, 	
	$\mathbf { F }=   \frac{F}{\omega} \,\text{mod}\,q,   \mathbf { G }= \frac{  G}{s_1 \omega} \,\text{mod}\,q$,  
	$\mathbf { R }= \frac{  R}{s_1 \omega} \,\text{mod}\,q$,
	$\mathbf { H }= \frac{  H}{s_2} \,\text{mod}\,q$,  $\mathbf {  J }=  \frac{J}{s_1 s_2} \,\text{mod}\,q$ and $\mathbf { S }= \frac{ S}{s_1 s_2} \,\text{mod}\,q$. We emphasize that re-encryption of the control input is not required in view of (\ref{fifth controller}) and (\ref{5.5 controller}).
	
	By Proposition 1 in \cite{kimcdc}, one has
$
	  \textbf{Dec}(\mathbf {  u  }(t)) = \tilde u (t) \,\text{mod}\,q,
$
in which $\tilde u (t) \,\text{mod}\,q$ is not necessarily equivalent to $\tilde u (t)$. 
 To restore $\tilde u (t)$ from $\tilde u (t) \,\text{mod}\,q$, the actuator operates the following algorithm \cite{kimcdc}, obtaining $\tilde 	u_a(t)$ and $u_a(t)$:
\begin{subequations}\label{15}
\begin{align}
	&\tilde 	u_a(t): =
	 \textbf{Dec}(\mathbf {  u }(t)) - \Big\lfloor \frac{\textbf{Dec}(\mathbf {  u }(t))  \!-\! \frac{\tilde u(t-1)}{\omega} + \frac{q}{2}}{q} \Big\rfloor  q   \\
	&u_a(t):=  s_1 s_2 l(t) \tilde u_a(t).	 
\end{align}
\end{subequations}

The following proposition presents a preliminary result whose proof is provided in Appendix. 
\begin{proposition}\label{proposition 1}
Suppose that Lemma \ref{lemma} holds. Consider the encrypted controller (\ref{5.5 controller}) and the the algorithm on the actuator side (\ref{15}).
If $q> 2\|[JC\,\,\, H]\|_\infty M$, then $\tilde u_a(t)= \tilde u(t)$ and hence $ u_a(t)=   u(t)$ with $u(t)$ in (\ref{5}), where $M$ is an upper bound of $\|\delta(t)\|$ with $\delta(t):=[\bar x_p ^T (t)- \bar x_p ^T (t-1)/\omega\,\,\,\,\bar x ^T  (t)- \bar x^T (t-1)/\omega]^T$.\qedp
\end{proposition}

\begin{remark}\label{remark 1}
We emphasize that the condition $\rho_c< s_F$ in Lemma \ref{lemma} is important for ensuring a finite $q$. Specifically, one needs $\rho_c< s_F$ to ensure $\rho(A_{\text{cl}}/\omega) <1$ (see (\ref{delta}) in Appendix), under which $\|\delta(t)\|$ is upper bounded. 
If $\rho _ c \ge s_F$, a sufficiently small $\omega\le s_F$ that scales $F/\omega$ into an integer matrix can lead to $\omega \le s_F \le \rho_c$, under which $\rho(A_{\text{cl}}/\omega)\ge 1$. Then, by (\ref{delta}), $\|\delta(t)\|$ can be unbounded. This implies that $\|\tilde u(t)  -  \tilde u(t-1)/\omega\|_\infty \le \|[JC\,\,\, H]\|_\infty \|\delta(t)\|$ can be unbounded. Consequently, one needs an infinitely large $q$ such that  $ \Big\lfloor \frac{\tilde u(t)  - \tilde u(t-1)/\omega +q/2}{q} \Big\rfloor = \mathbf 0$, in order to realize $\tilde u_a(t)=\tilde u(t)$ on the actuator side by (\ref{15}). In other words, there does not exist a finite $q$ such that  $ \Big\lfloor \frac{\tilde u(t)  - \tilde u(t-1)/\omega +q/2}{q} \Big\rfloor = \mathbf 0$. \qedp
\end{remark}

\begin{remark}\label{remark 2}
If the poles of the closed-loop system can be arbitrarily placed, then there always exist an $\omega$ satisfying Lemma \ref{lemma} and a finite $q$ satisfying Proposition \ref{proposition 1}. For example, one can let all the closed-loop poles be 0, i.e., $\rho_c=0$ in (\ref{rho_c}). Then, there must exist a sufficiently small $\omega$ satisfying (1) in Lemma \ref{lemma} because $\omega$ is lower bounded by $\rho_c = 0$. Importantly, such an $\omega$ can still ensure that $\delta(t)$ is upper bounded because $\rho(A_{\text{cl}}/\omega)=0$ in (\ref{delta}). Then finding a finite $q$ is straightforward by following the analysis after  (\ref{delta}).
However, if the controller (\ref{first controller}) is pre-given such that the closed-loop poles are pre-fixed (e.g., $\rho_c$ is ``large"), the results in Lemma \ref{lemma} and Proposition \ref{proposition 1} are not applicable. To deal with a pre-given controller, a new control scheme will be provided in the next subsection. \qedp

\end{remark}

 Actually, Lemma \ref{lemma} and Proposition \ref{proposition 1} are the generalized approaches of those in \cite{feng2024asymptotic}. In \cite{feng2024asymptotic}, because the closed-loop poles depending on $A+BK$ (feedback gain $K$) and $A-LC$ (observer gain $L$) can be arbitrarily placed, then there must exist a sufficiently small $\omega$ and a finite $q$.  

\subsection{Main result}




In the former subsection, we have shown that if  $\rho_c < s_F$, then re-encryption is not required. 
However, this condition would be very hard to be satisfied, for a pre-given controller.
In this subsection, we present the main result of the paper, which shows that it is possible to encrypt a controller without requiring $\rho_c < s_F$ and re-encryption.

\begin{assumption}
The pair $(A,C)$ in (\ref{process}) is observable. 
\end{assumption}

For the ease of conveying the fundamental idea of encrypted controller design, we start with the controller design in plaintexts. 
The new controller approximating the original one in (\ref{first controller}) has the following structure.  
\begin{align}\label{observer} 
\left\{\!\!
\begin{array}{l}
	x_o(t\!+\!1)  = A  x_o(t) \!+\! B u(t) \!+\! l(t)L   Q\left(\!\frac{y_p(t) - y_o(t)}{l(t)}\!\right) \\
  	y_o(t)= Cx_o(t)\\
	r_e(t+1)= r_e(t) + l(t)Q\left(\frac{r(t)- r_e(t)}{l(t)}\right)\\
	x_e(t+1)=Fx_e(t)+GC x_o(t) + R r_e(t)\\
u_e(t)=Hx_e(t)+JCx_o(t) + S r_e(t)
\end{array}
\right.
\end{align}
where $x_o(t)\in \mathbb{R}^{n\times n}$ is the state of the observer, $r_e(t)\in \mathbb R^{n_r}$, $x_e(t)\in \mathbb R^{n_x}$ and $u_e(t)\in \mathbb R^{w}$ are the estimates of $r(t)$, $x(t)$ and $u(t)$, respectively,  $L\in \mathbb{R}^{n\times v}$ is the observer gain, and $Q(\cdot)$ denotes a finite-capacity quantizer, whose quantization range will be provided later. By convention, we let $x_o(0)=\mathbf 0$. 
By defining $\bar x_o(t) =  x_o(t)/l(t)$, the observer dynamics can be written as
$\bar x_o(t+1) = \frac{A}{\omega} \bar x_o(t)+ \frac{B}{\omega}  \bar u(t)  +   \frac{L}{\omega} Q\left(\frac{y_p(t)- y_o(t)}{l(t)}\right)$. Similarly, by defining $\bar r_e(t):=\frac{r_e(t)}{l(t)}$ and $\bar r(t):=\frac{r(t)}{l(t)}$, we obtain $	\bar r_e(t+1)= \frac{\bar r_e(t)}{\omega} +  \frac{1}{\omega}Q\left(\frac{r(t)- r_e(t)}{l(t)}\right)$. 
The controller (\ref{observer}) can be converted to 
\begin{subequations}\label{sixth controller}
\begin{align}
	\bar x_o(t+1)&\!= \!\frac{A}{\omega} \bar x_o(t) \!+ \! \frac{s_2B}{\omega}  \tilde  u_e(t)  \!+\!   \frac{L}{\omega} Q\left(\!\frac{y_p(t) \!-\!  y_o  ^s  (t)}{l(t)}\!\right)\\
 \tilde	y_o(t)&= \frac{C}{s_1} \bar x_o(t)\\
	\bar r_e(t+1)&= \frac{\bar r_e(t)}{\omega} +  \frac{1}{\omega}Q\left(\frac{r(t)- r_e(t)}{l(t)}\right)\\
	\bar x_e(t+1)& =\frac{F}{\omega} \bar x_e(t)+\frac{GC}{\omega} \bar x_o(t) +  \frac{R}{\omega}  \bar r_e (t)    \\
	\tilde u_e(t)& =\frac{H}{s_2}  \bar x_e(t)+\frac{JC}{s_2}\bar x_o(t) +\frac{S}{s_2} \bar r_e(t)
 \end{align}
\end{subequations}
where $\bar x_e(t)= x_e(t)/l(t)$, $\tilde u_e(t)= u_e(t)/(s_2l(t))$, 
and $Q (\frac{y_p(t)-  y_o ^s (t)}{l(t)} )$ and $Q (\frac{r(t)- r_e(t)}{l(t)} )$ are integer vectors received from the sensors and reference provider, respectively. The variable $y_o ^s(t) = y_o(t)$ will be explained later. To apply (\ref{sixth controller}), the sensors and the reference provider should know $y_o ^s(t) = y_o(t)$ and $r_e(t)$, respectively. In our paper, we assume that the sensors are capable of receiving information. Specifically, the sensors receive the ciphertexts of $\tilde  y_o(t)$ soon after $t-1$.
Then, the sensors decrypt the ciphertexts of $\tilde  y_o(t)$ and generate $y_ o^s (t)$. As will be shown later in (\ref{sensor side}), $y_ o^s (t) =y_o (t)$ can be ensured. The reference provider operates a copy of $	r_e(t+1)= r_e(t) + l(t)Q (\frac{r(t)- r_e(t)}{l(t)} )$ locally and thus knows $r_e(t)$. 

We assume that the controller has a memory such that its historical data (of $t-1$ and $t-2$) is available at $t$. Thus, the controller transmits
\begin{subequations}\label{controller output}
\begin{align}
\!\! &\!\!\alpha(t):=	\bar x_o(t) - \frac{A}{\omega} \bar x_o(t-1) - \frac{ s_2 B}{ \omega} \tilde u_e(t-1)\\
	&\!\!\beta(t)\!: =\! \bar x_e(t) \!-\! \frac{F}{\omega} \bar  x_e(t\!-\!1) \!-\! \frac{GC}{\omega}\bar  x_o(t\!-\!1) \!-\! \frac{1}{\omega}  r_\beta(t\!-\!1) \\
	&\!\! \gamma(t):= \tilde u_e(t) - \frac{H}{s_2} \bar x_e(t) - \frac{JC}{ s_2} \bar  x_o(t) - \frac{1}{\omega}  r_ \gamma (t-1) 
\end{align}
\end{subequations}
to the actuator, in which 
\begin{subequations}\label{r}
\begin{align}
	& r _\beta(t-1)\! : =\!    \bar x_e(t-1) - \frac{F}{\omega}\bar  x_e(t-2) - \frac{GC}{\omega}\bar x_o(t-2) \\
	& r_ \gamma (t-1)\!: =\! \tilde   u_e(t-1)\!-\! \frac{H}{s_2} \bar  x_e(t-1)\! -\! \frac{JC}{ s_2} \bar x_o(t-1).
\end{align}
\end{subequations}

\begin{lemma}\label{lemma 2}
By the following steps, the matrices in (\ref{sixth controller}) and (\ref{r}) can be converted to integer matrices: 
\begin{itemize}
\item[1)] Because of $(A,C)$ being observable, select an $L$ such that $\rho(A-LC)=0$.

\item[2)] Select a sufficiently small $s_1$ and $s_2$ such that $\frac{C}{s_1}$, $\frac{H}{s_2}$, $\frac{JC}{s_2}$ and $\frac{S}{s_2}$ are integer matrices.

\item[3)] Select a sufficiently small $\omega$ such that $\frac{A}{\omega}$,  $\frac{s_2B}{\omega}$, $\frac{L}{\omega}$, $\frac{F}{\omega}$, $\frac{GC}{ \omega}$, $\frac{R}{\omega}$ and $\frac{1}{\omega}$ consist of integers. \qedp
\end{itemize}
\end{lemma}

The proof of Lemma \ref{lemma 2} is omitted. Moreover, in (\ref{sixth controller}) and (\ref{r}), all the states contain only integers for all $t$, which is implied by $\bar x_o(0)=\mathbf 0$ and $\bar x_e(0)=x_e(0)/l(0) =x(0)/l(0)\in \mathbb Z ^{n_x}$ under a small $l(0)$.
We then project (\ref{sixth controller})-(\ref{r}) over $\mathbb{Z}_q$ and obtain
\begin{align*} 
	&
	\begin{array}{l}
	\bar  x_o ^q (t\!+\!1)= \frac{A}{\omega} \bar x_o ^q(t)+ \frac{s_2B}{\omega}  \tilde  u^q(t)  +   \frac{L}{\omega} Q\left(\!\frac{y_p(t)- y ^s _o(t)}{l(t)}\!\right)  \,\text{mod}\, q\\
	 \tilde	y_o ^q (t)= \frac{C}{s_1} \bar x_o(t) \,\text{mod}\, q\\ 
	\bar r_e ^q (t+1)= \frac{\bar r_e ^q (t)}{\omega} +  \frac{1}{\omega}Q\left(\frac{r(t)- r_e(t)}{l(t)}\right)   \,\text{mod}\, q  \\
\bar x_e ^q (t+1) =\frac{F}{\omega} \bar x_e ^q (t)+\frac{GC}{\omega} \bar x_o ^q (t) +  \frac{R}{\omega}  \bar r_e ^q (t)    \,\text{mod}\, q  \\
\tilde u_e ^q (t) =\frac{H}{s_2}  \bar  x_e ^q (t)+\frac{JC}{s_2}\bar x_o^q(t) +\frac{S}{s_2} \bar r_e ^q (t)  \,\text{mod}\, q
	\end{array}
 \\
	&	\left\{\!\!\!	
	\begin{array}{l}
		\alpha^q(t):=	\bar x_o^q(t) - \frac{A}{\omega} \bar x_o ^q (t-1) - \frac{ s_2 B}{ \omega} \tilde u_e ^q (t-1) \,\text{mod}\, q\\
		\beta^q(t)\!: =\! \bar  x_e^q(t) \!-\! \frac{F}{\omega} \bar  x_e^q(t\!-\!1) \!-\! \frac{GC}{\omega}\bar  x_o^q(t\!-\!1) \!-\! \frac{r_\beta(t\!-\!1)}{\omega}   \,\text{mod}\, q\\
		\gamma^q(t):= \tilde u_e^q(t) - \frac{H}{s_2} \bar x_e^q(t) - \frac{JC}{  s_2} \bar  x_o^q(t) - \frac{r_ \gamma (t-1) }{\omega}   \,\text{mod}\, q 
	\end{array}
	\right.
\end{align*}
with $\bar x_o^q (0)= \bar x_o (0)$ mod $q$,  $\bar r_e^q (0)= \bar r_e (0)$ mod $q$ and  $\bar x_e ^q (0)= \bar x_e (0)$ mod $q$. 
Based on the projected dynamics over $\mathbb{Z}_q$, we present the following controller over homomorphically encrypted data
\begin{subequations}\label{seventh controller}
\begin{align}
	& \mathbf {   x_o  }(t+1)= \mathbf {   A_o \cdot  x_o  }(t)  \oplus \mathbf {   B_o \cdot  u  }(t) \nonumber\\
	&\quad \quad \quad   \oplus  \mathbf {   L_o    } \cdot \textbf{Enc}\left( Q\left(\!\frac{y_p(t)-  y_o ^ s(t)}{l(t)}\!\right) \right) \,\text{mod}\,q  )     \\
	&\mathbf{y_o}(t)=\mathbf C \cdot \mathbf {   x_o  }(t)\\
	&\mathbf{r }  (t+1)\!=\! \boldsymbol{\omega} \! \cdot \! \mathbf{r } (t) \!\oplus\!  \boldsymbol{\omega} \cdot  \textbf{Enc}\left(\! Q\left(\! \frac{r(t) \!-\! r_e(t)}{l(t)}\!\right)  \text{mod}\,q \! \right)    \\
	&	\mathbf {   x  }(t+1)= \mathbf {   F \cdot  x  }(t) \oplus \mathbf {   G  \cdot x_o  }(t)   \oplus   \mathbf {   R \cdot  r  }(t)\\
	&\mathbf { u  }(t)= \mathbf {   H  \cdot  x  }(t) \oplus \mathbf {   J \cdot  x_o  }(t) \oplus   \mathbf {   S  \cdot  r  }(t)
\end{align}
\end{subequations}
in which $\mathbf {   x_o  }(t), \mathbf{y_o}(t), \mathbf{r }  (t) , 	\mathbf {   x  }(t)$ and $\mathbf { u  }(t)$ are ciphertexts, and $ \mathbf { A_o } :=   \frac{A}{\omega} \,\text{mod}\,q$,    $\mathbf { B_o }:=   \frac{ s_2 B}{\omega} \,\text{mod}\,q$, $\mathbf { L_o }:=   \frac{L}{\omega} \,\text{mod}\,q$, $\mathbf C:= \frac{C}{s_1} \, \text{mod}\, q,$
$\mathbf { F }=   \frac{F}{\omega} \,\text{mod}\,q,   \mathbf { G }= \frac{  GC}{  \omega} \,\text{mod}\,q,  
\mathbf { R }= \frac{  R}{  \omega} \,\text{mod}\,q, \mathbf { H }= \frac{  H}{s_2} \,\text{mod}\,q,  \mathbf {  J }=  \frac{JC}{  s_2} \,\text{mod}\,q,  \mathbf { S }= \frac{ S}{ s_2} \,\text{mod}\,q$, $\boldsymbol{\omega }:=  \frac{I_{n_r}}{\omega}\,\text{mod}\, q$,  
 $\mathbf {   x_o  }(0) =\textbf{Enc}(\bar x_o ^q(0))$, $\mathbf {   r  }(0) =\textbf{Enc}(\bar r_e ^q(0))$, and $\mathbf {  x(0) } = \textbf{Enc}(\bar  x_e^q(0))$. In (\ref{seventh controller}), one can see that re-encryption of the control input is not required.

The encrypted controller transmits $\boldsymbol{\alpha}(t)$, $\boldsymbol{\beta}(t)$ and $\boldsymbol{\gamma}(t)$ (ciphertexts) to the actuator as 
\begin{subequations}\label{output}
\begin{align}
	&\boldsymbol{\alpha}(t):= \mathbf {   x_o  }(t) \oplus \mathbf A' \cdot\mathbf {   x_o  }(t-1) \oplus \mathbf B' \cdot \mathbf u(t-1)\\
	&\boldsymbol{\beta}(t):= \mathbf {   x  }(t) \oplus \mathbf F'\cdot \mathbf {   x  }(t-1) \oplus \mathbf G' \cdot \mathbf x_o(t-1) \nonumber\\
	&\quad\quad \quad \oplus \boldsymbol{\omega'} \cdot \boldsymbol{r_\beta}(t-1)\\
	&\boldsymbol{\gamma}(t): = \mathbf u(t) \oplus \mathbf H' \cdot \mathbf x(t) \oplus \mathbf J' \cdot \mathbf x_o(t) \oplus   \boldsymbol{\omega'} \cdot \boldsymbol{r_\gamma}(t-1)
\end{align}
\end{subequations}
in which 
\begin{subequations}\label{r_beta}
\begin{align}
	&\boldsymbol{r_\beta}(t\!-\!1): = \mathbf x (t \!-\!1) \oplus \mathbf F' \cdot \mathbf x(t-2)\oplus \mathbf G' \cdot  \mathbf x_o(t-2)\\
	&\boldsymbol{r_\gamma}(t\!-\!1): = \mathbf u(t\!-\!1) \oplus \mathbf H' \cdot \mathbf x(t\!-\!1) \oplus \mathbf J' \cdot  \mathbf x_o(t-1)
\end{align}
\end{subequations}
$\mathbf {A'} : = -\frac{A}{\omega} \,\text{mod}\, q$,
$\mathbf {B'}: = -\frac{ s_2 B}{\omega} \,\text{mod}\, q$, 
$\mathbf{F'}: = -\frac{F}{\omega} \,\text{mod}\, q$, 
$\mathbf{G'}: = -\frac{GC}{  \omega}\,\text{mod}\, q$
$\mathbf {H'}: = - \frac{H}{s_2}\,\text{mod}\, q$,
$\mathbf {J'}: = - \frac{JC}{  s_2}\,\text{mod}\, q$, 
and $\boldsymbol{\omega'}:= -  \frac{I_{n_r}}{\omega}\,\text{mod}\, q $.

On the actuator side, it first computes
\begin{subequations}\label{26}
\begin{align}
&\alpha^a(t): = \textbf{Dec}(\boldsymbol{\alpha}(t)) - \Big\lfloor \frac{\textbf{Dec}(\boldsymbol{\alpha}(t))+ \frac{q}{2}}{q} \Big \rfloor q \\
&\beta^a(t): = \textbf{Dec}(\boldsymbol{\beta}(t)) - \Big\lfloor \frac{\textbf{Dec}(\boldsymbol{\beta}(t))+ \frac{q}{2}}{q} \Big \rfloor q \\
&\gamma^a(t): = \textbf{Dec}(\boldsymbol{\gamma}(t)) - \Big\lfloor \frac{\textbf{Dec}(\boldsymbol{\gamma}(t))+ \frac{q}{2}}{q} \Big \rfloor q 
\end{align}
\end{subequations}
and then calculates
\begin{subequations}\label{27}
\begin{align}
	&x_o ^a (t):=	l(t) \alpha^a(t) + Ax_o ^a (t-1) +  B u^a (t-1) \\
	&x^a (t): = l(t) \beta^a(t) + Fx^a(t-1) +GC x_o ^a (t-1) \nonumber\\
	&\quad\quad + (x^a (t-1)- Fx^a (t-2) -GC x_o ^a (t-2))  \\
	&u^a(t): =  s_2 l(t) \gamma^a(t)  + Hx^a(t)+ JC x^a _o(t)  \nonumber\\
	& \quad\quad+ (u^a(t-1)- H x^a(t-1)-JCx_o ^a (t-1)) 
\end{align}
\end{subequations}
where we assume that the actuator has a memory such that its historical data ($t-1$ and $t-2$) is available at $t$.

In the following, we explain the data flow in the closed-loop system. 
We assume that the communication is free of delay. Soon after $t-1$, $ \mathbf {   x_o  }(t)$ is calculated by the controller as $ \mathbf {   x_o  }(t)= \mathbf {   A_o \cdot  x_o  }(t-1)  \oplus \mathbf {   B_o \cdot  u  }(t-1) \oplus  \mathbf {   L_o    } \cdot \textbf{Enc}\left( Q\left(\frac{y_p(t-1)-  y_o ^s(t-1)}{l(t-1)}\right) \right) \,\text{mod}\,q  )$. Then, the controller generates $\mathbf{y_o}(t)=\mathbf C \cdot \mathbf {   x_o  }(t)$  and sends it to the sensors. When the sensors receive $\mathbf{y_o}(t)$, it computes
\begin{align}\label{sensor side}
	y_o^s(t)=s_1l(t)\left(\textbf{Dec}(\mathbf{y_o}(t)) - \Big\lfloor \frac{\textbf{Dec}(\mathbf{y_o}(t))  \!-\!  \frac{\bar y_p(t) }{s_1} + \frac{q}{2}}{q} \Big\rfloor  q\right) 
\end{align} 
in which $q$ will be provided in Theorem 1. Subsequently, 
the sensors transmit ciphertexts $\textbf{Enc} ( Q ( \frac{y_p(t)-  y_o ^s(t)}{l(t)} )  ) \,\text{mod}\,q  ) $ to the controller. Meanwhile, the reference provider sends $ \textbf{Enc}( Q( \frac{r(t)- r_e(t)}{l(t)})  \text{mod}\,q  )   $ to the controller.
Recall that $\mathbf {   x_o  }(t-1)$, $\mathbf {   x  }(t-1)$, $\mathbf {   u  }(t-1)$, $\mathbf {   x_o  }(t-2)$ and $\mathbf {   x  }(t-2)$ are stored in the memory of the controller, then the controller is able to calculate $\boldsymbol{\alpha}(t)$, $\boldsymbol{\beta}(t)$ and $\boldsymbol{\beta}(t)$ by (\ref{output}), and sends them to the actuator. The actuator decrypts them by $\textbf{Dec}(\cdot)$ and calculates $\alpha^a(t)$, $\beta^a(t)$ and $\gamma^a(t)$ as in (\ref{26}). Recall that $x_o ^ a(t-1)$, $u^a (t-1)$, $x^a (t-1)$, $x^a (t-2)$ are stored in the memory of the actuator. Eventually, the actuator computes $x_o^a(t)$, $x^a(t)$ and $u^a(t)$ as in (\ref{27}).

Please note that re-encryption of the control inputs (e.g., in \cite{kimtac1} and in Section II-B of our paper) is for preventing overflow problems in the cryptosystems, while the sensors decrypt $\mathbf{y_o}(t)$ in (\ref{sensor side}) and encrypt $ Q ( \frac{y_p(t)-  y_o ^s(t)}{l(t)} ) $ is for preventing quantizer saturation. In our paper, by re-encryption, we refer to re-encryption of the control inputs. 
 
We are ready to present the main result of the paper.

\begin{theorem}\label{Theorem 1}
 Consider the encrypted controller (\ref{seventh controller})-(\ref{r_beta}), and the algorithm embedded in the actuator (\ref{26})-(\ref{27}). 
 If 
 \begin{align}\label{q result}
 q>  &\, \max\left\{   
  \frac{2C_e\|[LC\,\, L]\|_\infty}{\omega},   \frac{\|[R/\omega\,\, -R]\|_\infty}{\omega^2}, \right. \nonumber\\
  &\quad\quad \quad\left. \frac{\|[S/\omega\,\, -S]\|_\infty}{s_2\omega}, 2\|C/s_1\|_\infty C_e  \right\} 
 \end{align}
then one has
\begin{itemize}

\item [1)] $x_o^a(t)= x_o(t)$, $x^a(t)= x_e(t)$, $y_o^s(t) = y_o(t)$, $u^a(t)= u_e(t)$, and $\lim_{t\to\infty}u_e(t) = u(t)$, with $u(t)$ in (\ref{first controller}) and $u_e(t)$ in (\ref{observer}). 

\item [2)] The process is asymptotically stable. The quantizer $Q(\cdot)$ is not saturated if $(2R+1)/2 >\max\{\|C\|_\infty C_e, \frac{1}{2\omega}\}$. 
\end{itemize}
In (\ref{q result}), $s_2$ and $\omega$ are selected as in Lemma \ref{lemma 2}, $C_e$ is an upper bound of $\|\bar e _o(t)\|$ discussed after (\ref{e_o}). 
\end{theorem}

\textbf{Proof.} 
The proof will be conducted by two steps. 

\textbf{Step 1.} In this step, we will show that $\alpha^a(t)=\alpha(t)$, $\beta^a(t)=\beta(t)$ and $\gamma^a(t)=\gamma(t)$. 
We use an induction type of analysis. That is, if $y_o ^s (k)= y_o(k)$, $\alpha^a(k)=\alpha(k)$, $\beta^a(k)=\beta(k)$, $\gamma^a(k)=\gamma(k)$ and the quantizer is not saturated for $k=0, \cdots, t-1$, then one must obtain $y_o ^s (t)= y_o(t)$, $\alpha^a(t)=\alpha(t)$, $\beta^a(t)=\beta(t)$, $\gamma^a(t)=\gamma(t)$ and the quantizer is not saturated at $t$.

In view of $\textbf{Dec}(\boldsymbol{\alpha}(t))= \alpha(t) \, \text{mod}\,q$, one can obtain $\alpha^a(t)  = \alpha(t) - \Big\lfloor \frac{\alpha(t)+ \frac{q}{2}}{q} \Big \rfloor q$ by (\ref{26}a). Then, we are interested if $\alpha(t)$ can be upper bounded by $\frac{q}{2}$. 
 Substituting $\bar x_o(t)= \frac{A}{\omega} \bar x_o(t-1)+ \frac{s_2B}{\omega}  \tilde  u_e(t-1)  +   \frac{L}{\omega} Q\left(\frac{y_p(t-1)-  y_o ^s(t-1)}{l(t-1)}\right)$ into (\ref{controller output}a), one can obtain $\alpha(t)=  \frac{L}{\omega}  Q (C \bar e_o(t-1) )$, in which $\bar e_o(t) := \bar x_p(t) - \bar x_o(t)$. To verify that $\bar e_o(t)$ is upper bounded, we present its dynamics
\begin{align}\label{e_o}
\bar e_o(t+1)= \frac{A-LC}{\omega}\bar e_o(t) - \frac{L}{\omega} \bar e_{\bar y}(t)
\end{align}
in which $\bar e_{\bar y}(t):= Q(C  \bar e_{ o}(t))- C \bar e_{ o}(t)$ and $\rho(\frac{A-LC}{\omega})=0$ because $\rho(A-LC)=0$. Moreover, the quantization error satisfies $\|\bar e_{\bar y}(t)\|_\infty \le \frac{1}{2}$ because it is not saturated by hypothesis. Then, it is simple to verify that there must exist a $C_{e} \in \mathbb{R}_{\ge \frac{1}{2}}$ such that $\|\bar e_o(t)\| \le C_e $. One can obtain  
\begin{align}
\|\alpha(t)\|_\infty\le \frac{1}{\omega}
\left\|\left[LC\,\, L\right]\right\|_\infty  \| [\bar e_{o} ^T (t)\,\, \bar  e_{\bar y}^T(t)]^T\|_\infty
< \frac{q}{2}
\end{align}
 and hence $\lfloor \frac{\alpha(t)+ q/2}{q}  \rfloor = \mathbf 0$, which implies $\alpha^a(t)=\alpha(t)$.

Similarly, to show $\beta^a(t)=\beta(t)$, one essentially needs to show $\|\beta(t)\|_\infty < \frac{q}{2}$. Substituting the dynamics of $	\bar  x_e(t)  $ and $	\bar  x_e(t-1)  $  into (\ref{controller output}b) and (\ref{r}a), respectively, 
one has 
$
\beta(t) =  \frac{R}{\omega} \bar r_e (t-1) - \frac{1}{\omega} \gamma_\beta (t-1) =  \frac{R}{\omega} \bar r_e (t-1)- \frac{1}{\omega} \frac{R}{\omega} \bar r_e (t-2)= [\frac{R}{\omega^2}\,\,-\frac{R}{\omega}] [\bar e_r ^T(t-2)\,\,\bar e_r ^T(t-1)]^T
$, in which $\bar e_r(t) := \bar r(t) - \bar r_e(t)$ with $\|\bar e_r(t) \|_\infty \le \frac{1}{2\omega}$  \cite{feng2024asymptotic}.
Then, it is clear that 
\begin{align}
\|\beta(t)\|_\infty  \le   \left\| \left[\frac{R}{\omega^2}\,\,-\frac{R}{\omega}\right]  \right\|_\infty   \frac{1}{2\omega} <\frac{ q}{2},
\end{align}
which implies $\lfloor \frac{\beta(t)+ q/2}{q}  \rfloor = \mathbf 0$. Therefore, one has $\beta^a(t)=  \beta(t) -  \lfloor \frac{\beta(t)+ q/2}{q}   \rfloor q  =\beta (t)$.

To show $\gamma^a(t)=\gamma(t)$, one needs to prove $\|\gamma(t)\|_\infty < \frac{q}{2}$. Substituting the equations of $\tilde u_e(t)$ and  $\tilde u_e(t-1)$ into (\ref{controller output}c) and (\ref{r}b), respectively, one has $\gamma(t)= \frac{S}{s_2}\bar r_e(t) - \frac{S}{\omega s_2} \bar r_e(t-1)=[\frac{S}{s_2\omega}\,\,-\frac{S}{s_2}] [\bar e_r ^T(t-1)\,\,\bar e_r ^T(t)]^T$.
Similarly, one can verify that
\begin{align}
& \|\gamma(t)\|_\infty  \le   \left\| \left[\frac{S}{s_2\omega}\,\,-\frac{S}{s_2}\right]  \right\|_\infty   \frac{1}{2\omega}< q/2,
\end{align}
which implies $\lfloor \frac{\gamma(t)+ \frac{q}{2}}{q}  \rfloor = \mathbf 0$. Therefore, one has $\gamma^a(t)=  \gamma(t) -  \lfloor \frac{\gamma(t)+ q/2}{q}   \rfloor q  =\gamma (t)$.

By $\alpha^a(t)=\alpha(t)$ and substituting $l(t)\alpha(t)$ (in (\ref{controller output}a) ) into $x_o ^a(t)$ (in (\ref{27}a)), one can obtain $x_o^a(t)= x_o(t)- A x_o(t-1)-Bu_e(t-1)+A x_o(t-1)+Bu_e(t-1)= x_o(t) $. Second, by $\beta^a(t)=\beta(t)$ and substituting $ l(t) \beta(t) $ into (\ref{27}b), one can obtain $x  ^a(t)= x_e (t)$.  Similarly, substituting $s_2 l(t) \gamma(t)$ into (\ref{27}c), one can obtain $u^a(t)= u_e(t)$.

On the sensor side, (\ref{sensor side}) is equivalent to
\begin{align}\label{31}
	y_o^s(t)&=s_1l(t)\left( \frac{\bar y_o(t)}{s_1}  - \Big\lfloor \frac{\frac{\bar y_o(t)}{s_1}    \!-\!  \frac{\bar y_p(t) }{s_1} + \frac{q}{2}}{q} \Big\rfloor  q\right)  \nonumber\\
	&= s_1 l(t)  \bar y_o(t)/s_1 = y_o(t)
\end{align} 
in which $\|\frac{\bar y_o(t)}{s_1}    - \frac{\bar y_p(t) }{s_1}\|_\infty =\| \frac{C( \bar x_o(t) - \bar x_p(t))}{s_1}   \|_\infty  = \|\frac{C}{s_1} \bar e_o(t)\| _\infty \le  \|C/s_1\|_\infty C_e  < q/2$, and hence $\lfloor \frac{\frac{\bar y_o(t)}{s_1}    \!-\!  \frac{\bar y_p(t) }{s_1} + \frac{q}{2}}{q}  \rfloor= \mathbf 0$. Then, one can see that if $y_o^s(t-1) = y_o(t-1)$, then $	y_o^s(t)=y_o(t)$. 
 
 To show that the quantizer $Q(\cdot)$ is free from saturation at $t$, it is sufficient to show $\|\frac{y_p(t)- y_ o  ^s (t)}{l(t)}\|_\infty =\|\frac{y_p(t)- y_o(t)}{l(t)}\|_\infty= \|C \bar e_o (t)\|_\infty \le \|C\|_\infty C_e < \frac{2R+1}{2} $ and $\|\frac{r(t)- r_e(t)}{l(t)}\|_\infty = \|  \bar e_r (t)\|_\infty \le \frac{1}{2\omega} < \frac{2R+1}{2} $.  

\textbf{Step 2.} The dynamics of $\bar x_e(t)$ and $u_e(t)$ in (\ref{observer}) is equivalent to:
\begin{align}\label{34}
	\left\{
	\begin{array}{l}
		x_e(t+1)=Fx_e(t) + Gy(t) + Rr(t)+ e_1(t)\\
		u_e(t)=Hx_e(t)+Jy(t) + Sr(t)+  e_2(t)
	\end{array}
	\right.
\end{align}
in which 
$e_1(t):=-l(t)(GC \bar e_{o}(t) + R \bar e_{r}(t))$ and $e_2(t)=-l(t)(JC \bar e_{o}(t) -S \bar e_{r}(t))$. Because of $l(t)\to 0$, and $\bar e_{o}(t)$ and $\bar e_{r}(t)$ are upper bounded, one has $e_1(t)$ and $e_2(t)$ converge to zero asymptotically. Therefore, one can verify that $x(t)-x_e(t)\to0$, $u(t)-u_e(t)\to 0$, $u(t)-u^a(t)\to 0$ asymptotically under Assumption \ref{ass 1}. Then, it is straightforward that the process is asymptotically stable. 
\qedp

\begin{remark}
One can see that $\omega>\rho_c$ (in Lemma \ref{lemma}) is not required in Lemma \ref{lemma 2}. Then, there always exists a sufficiently small $\omega$ such that the controller (\ref{sixth controller}) has only integer matrices.
		 In principle, $\omega$ should be greater than $\rho(A-LC)$, under which $\bar e_o(t)$ in (\ref{e_o}) does not diverge. However, thanks to $\rho(A-LC)=0$ by observer design, one is able to let $\omega>\rho(A-LC)=0$ be arbitrarily small. If one does not implement the observer, $\omega$ will directly interact with the pre-given $A_{\text{cl}}$ (see (\ref{delta})) instead of $A-LC$, under which there does not exist a finite $q$ preventing overflow, or overflow will occur under any finite $q$ (see Remarks \ref{remark 1} and \ref{remark 2}).  
	\qedp 
\end{remark}

\begin{remark}
In (\ref{r}), $r_\beta(t-1)$ in $\beta(t)$ and $r_\gamma(t-1)$ in $\gamma(t)$ are necessary for ensuring a finite $q$. In the proof of Theorem \ref{Theorem 1}, we have shown that $\beta(t) =   \frac{R}{\omega} \bar r_e (t-1) - \frac{1}{\omega} \gamma_\beta (t-1) $. If one removes $r_\beta(t-1)$, $\beta(t)$ diverges due to $\lim_{t \to \infty} \bar r_e (t)= \infty $. Then, one cannot find a finite $q$ to realize $\beta^a(t)= \beta(t)$ by (\ref{26}b). Implementing $r_\gamma(t-1)$ in $\gamma(t)$ follows the similar reason. 
In a special case, if $S=0$ in the original controller (\ref{first controller}), then the controller only needs to transmit $\boldsymbol{\alpha}(t)$ and $\boldsymbol{\beta}(t)$ in (\ref{output}) to the actuator, under which $x_o(t)$ and $x_e(t)$ are restored by (\ref{26}) on the actuator side, respectively. 
\qedp 
\end{remark}

\begin{remark}
In the following, we discuss the necessities of transmitting $\boldsymbol{\alpha}(t)$, $\boldsymbol{\beta}(t)$ and $\boldsymbol{\gamma}(t)$ at each $t$. Suppose that the controller transmits only $\boldsymbol{\gamma}(t)$ to the actuator at each $t$. This implies that $x^a(t)$, $x^a_o(t)$, $x^a(t-1)$ and $x^a_o(t-1)$ are not available to the actuator for all $t$. In this case, it is very difficult to reconstruct $u^a(t)$ from only (\ref{27}c). If the actuator has access to $\boldsymbol{\alpha}(t)$ and $\boldsymbol{\beta}(t)$ ($S\ne 0$), but not $\boldsymbol{\gamma}(t)$, it is clear that $u^a(t) = Hx^a(t)+ JC x^a _o(t) + (u^a(t-1)- H x^a(t-1)-JCx_o ^a (t-1))$ by (\ref{27}c) could not match $u_e(t)$. 
In our paper, the controller does not transmit $\mathbf {   x_o  }(t)$, $\mathbf {   x }(t)$ and $\mathbf { u  }(t)$ to the actuator because their plaintexts ($\bar x_o(t)$, $\bar x_e(t)$ and $\tilde u(t)$) diverge, and hence it is not possible to restore them on the actuator side under a finite $q$ by (\ref{26}). For instance, after some $t$, one must have $\bar x_o(t) \ne \textbf{Dec}(\mathbf{x_o}(t)) - \Big\lfloor \frac{\textbf{Dec}(\mathbf{x_o}(t) )+ \frac{q}{2}}{q} \Big \rfloor q$. 
\qedp  
\end{remark}

\begin{table} 
	\caption{Communication and computation overheads}
	\begin{tabular}{ p{2.4cm}||p{2.5cm}|p{2.5cm}  }
		\hline
		& with re-encryption \newline in Section II-B&without re-encryption \newline in Section III-B\\
		\hline
		Communication
		\newline  overheads between controller \& actuator   & $\mathbf {  u  }(t) \in \mathcal C ^w$ 
		\newline \textbf{Enc}($Q(s_cs_u u_c(t))$  $\in \mathcal C ^{w}$   &$\boldsymbol{\alpha}(t)  \in \mathcal C ^{n}$
		$\newline \boldsymbol{\beta}(t) \in \mathcal C ^{n_x}$
		$\newline  \boldsymbol{\gamma}(t) \in \mathcal C ^{w}$ \\
		\hline
		Computation  \newline overheads of actuator &   \textbf{Dec}($\mathbf {  u  }(t))\in\mathbb Z ^{w}$
		\newline \textbf{Enc}($Q(s_cs_u u_c(t))$  $\in \mathcal C ^{w}$
		&  \textbf{Dec}($\boldsymbol{\alpha}(t)$) $\in\mathbb Z ^{n}$ \newline \textbf{Dec}($\boldsymbol{\beta}(t)$) $\in\mathbb Z ^{n_x}$  \newline \textbf{Dec}($\boldsymbol{\gamma}(t)$)  $\in\mathbb Z ^{w}$ \\
		\hline
	\end{tabular}
\end{table}

\textbf{Discussion on communication and computation overheads:} 
In the following, we compare the communication and computation overheads between the method utilizing re-encryption in Section II-B and the method without re-encryption in Section III-B. Specifically, the communication overheads refer to that taking place between the controller and actuator, and the computation overheads refer to that carried out by the actuator for encryption and decryption. The overheads of processing plaintexts will be left for future research.  

In view of Table I, the method utilizing re-encryption should exchange ciphertexts with size $\mathcal{C}^{2w }$ between the controller and actuator, while the method without re-encryption requires that the controller should transmit ciphertexts with size $\mathcal{C}^{w+n_x+n}$ to the controller. Therefore, the method without re-encryption proposed in this paper induces more communication overheads of size $\mathcal{C}^{n+n_x -w}$. 

For the computation overheads, it depends on the dimension of the inputs. First, note that the two methods  should process  \textbf{Dec}($\mathbf {  u  }(t))\in\mathbb Z ^{w}$ and \textbf{Dec}($\boldsymbol{\gamma}(t)$)  $\in\mathbb Z ^{w}$, respectively. Since they have the same dimensions, we assume that the computation overheads  of them are the same under the same key length. Then, the computation overheads  for processing \textbf{Enc}($Q(s_cs_u u_c(t)))\in \mathcal C ^{w}$ in re-encryption based methods, and for \textbf{Dec}($\boldsymbol{\beta}(t)$) $\in\mathbb Z ^{n_x}$ and \textbf{Dec}($\boldsymbol{\gamma}(t)$)  $\in\mathbb Z ^{w}$ in the method proposed in this paper are the key parts. We compare the computation time to inspect the computation overheads.
The papers \cite{darup2021encrypted} and \cite{farokhi2017secure} provide the computation time for encryption and decryption over the Paillier cryptosystem in the simulation examples. The computation time majorly depends on the key length. While under the same key length, the time for a single encryption is approximately 3-4 times of that for a single decryption, by the experimental data in \cite{darup2021encrypted} and \cite{farokhi2017secure}. Then, one can roughly estimate that if $n_x+n = (3\sim4)w$, the time for processing \textbf{Dec}($\boldsymbol{\beta}(t)$) $\in\mathbb Z ^{n_x}$ and  \textbf{Dec}($\boldsymbol{\gamma}(t)$)  $\in\mathbb Z ^{w}$ would be similar to that for processing \textbf{Enc}($Q(s_cs_u u_c(t))$  $\in \mathcal C ^{w}$. 
If $(3\sim4)w < n_x+n$, e.g., a single input control system ($w=1$), then the method utilizing re-encryption would consume less time because the actuator only needs to re-encrypt $w$ (few) states. If $n_x+n < (3\sim4)w$, then the method without re-encryption would be competitive. This is because if $w$ is ``large", re-encrypting the control inputs would need considerable computation. 

\section{Simulation}
In this section, we conduct simulation to verify the results of this paper. We consider the simulation example in \cite{murguia2020secure}, in which the authors discretize the batch reactor model in \cite{walsh2002stability} with sampling time 0.1s. Specifically, the discretized batch reactor model follows
\begin{align}
& A=
\begin{bmatrix}
1.18 &0 &0.51 & -0.4\\
-0.05&  0.66& -0.01 &0.06\\
0.08 & 0.34 & 0.56& 0.38\\
0 & 0.34 & 0.09& 0.85]
\end{bmatrix}
, B
=
\begin{bmatrix}
0 \\
0.47 \\
0.21 \\
0.21
\end{bmatrix}\nonumber\\
&C=
\begin{bmatrix}
1 & 0& 1 &-1\\
0&  1 & 0 &0
\end{bmatrix} \nonumber
\end{align}
and the matrices of the controller provided in \cite{murguia2020secure} follow
\begin{align}
&F= 
\begin{bmatrix}
0.26 &-0.03& -0.29 &0.31\\
-0.32 &1.24 &1.4 &-3.05\\
-0.45 &0.02& 0.87& -0.75\\
-0.05& -0.04& 0.72 &-0.51
\end{bmatrix}\nonumber\\
&G=
\begin{bmatrix}
-0.52 &-0.03\\
5.46& 1.25\\
 2.32 &-0.01\\
 2.28& -0.08
\end{bmatrix}, 
\begin{array}{l}
H=[1.02\,\, -2.65\,\, -2.65 \,\,6.28]\\
J=[-11.3 \,\,-4.09].
\end{array}
\nonumber
\end{align}
We let $R=   I _4$ and $S=[1\,\, 1\,\, 1\,\, 1].$
Given the process and controller above,  the closed-loop system without quantization and encryption is stable due to $\rho_c=0.8655$ defined in (\ref{rho_c}).

In light of $F$ above, $s_F=0.01$ is able to scale $F/s_F$ into an integer matrix. Because of $\rho_c > s_F$, in view of Lemma \ref{lemma}, one can infer that there does not exist an $\omega$ such that (\ref{fourth controller}) consists of integer matrices only. Thus, one needs to follow the scheme in Subsection III-B to design an encrypted controller. 
By the steps in Lemma \ref{lemma 2}, we first select
\begin{align}
L=\begin{bmatrix}
   2.0879 &    0.0705\\
-0.0024 &   1.4954\\
1.2623 &  15.4110\\
1.5956 &  14.5017
\end{bmatrix}
\end{align}
as the observer gain such that $\rho(A-LC)= 9 \times 10^{-7}$. Then, we select $s_1=1$, $s_2=0.01$ and $\omega=0.0001$ such that (\ref{sixth controller})-(\ref{r}) consist of only integer matrices. In the simulation, the reference is selected randomly as $r(t)=[1.1\,\, 5.2\,\, 3.5\,\, 6.7]^T$ for all $t$.

\begin{figure}[t]
	\begin{center}
		\includegraphics[width=0.44 \textwidth]{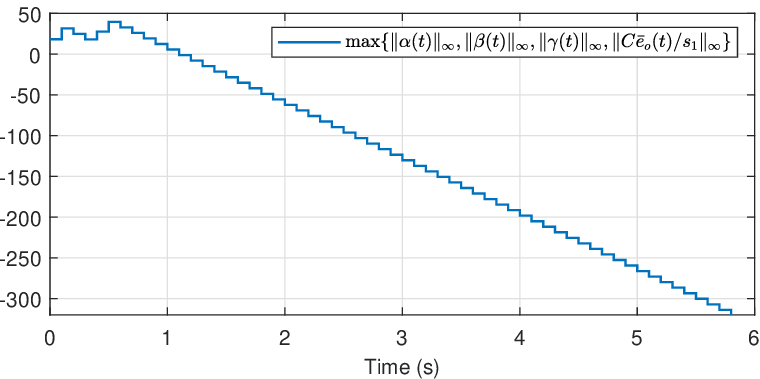}  \\
		\linespread{1}\caption{Time responses of $\log _2 \max\{\|\alpha(t)\|_\infty, \|\beta(t)\|_\infty, \|\gamma(t)\|_\infty,  \\
			\| \frac{C \bar e_o(t)}{s_1}\|_\infty \}$. } \label{Fig2}
	\end{center}
\end{figure}

Then, we attempt to find $q$. By the theoretical result (\ref{q result}), it is suggested that $q$ should be larger than $3.2508\times 10^{18}= 2^{ 61.4955}$. Then, selecting $q\ge2^{62}$ suffices the condition. 
Subsequently, by simulation, we also look for a $q$. 
According to Steps 1-3 in the proof of Theorem \ref{Theorem 1}, $q$ should be larger than $2 \max\{\|\alpha(t)\|_\infty, \|\beta(t)\|_\infty, \|\gamma(t)\|_\infty, \|\frac{C \bar e_o(t)}{s_1}\|_\infty \}$. The time responses of $\max\{\|\alpha(t)\|_\infty, \|\beta(t)\|_\infty, \|\gamma(t)\|_\infty , \|\frac{C \bar e_o(t)}{s_1}\|_\infty \}$ is presented in Fig. \ref{Fig2}, which is upper bounded by $2^{39.5}$. Then, selecting $q \ge  2^{41}$ is sufficient. 
Comparing the theoretical result $q\ge2^{62}$ and the simulation result $q\ge  2^{41}$, one can see that our theoretical result obtained by (\ref{q result}) is conservative. The conservativeness is caused by the worst-case type of analysis, the utilization of $\|\cdot\|_\infty$ and $\max\{\cdot\}$. 

\begin{figure}[t]
	\begin{center}
	\includegraphics[width=0.4 \textwidth]{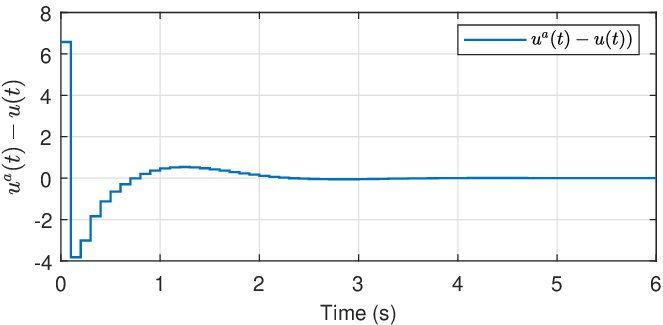}  \\
		\linespread{1}\caption{Time resporeses of $u^a(t)-u(t)$.} \label{Fig1}
	\end{center}
\end{figure}

In Fig. \ref{Fig1}, we show the discrepancy between $u^a(t)$ (the control input restored by the actuator under quantization and HE) and $u(t)$ (the control input of the original controller (\ref{first controller}) without quantization and HE). One can see that the actuator is able to restore the original control input asymptotically. At last, we mention that the quantizer is free of saturation under $(2R+1)/2 > 1.3594\cdot10^{13}$ and the process is asymptotically stable due to $u^a(t)-u(t)\to 0$. The plots are omitted.


\section{Conclusions}

This paper proposed two re-encryption free schemes to encrypt a pre-given control system with HE. The overflow problem under a finite modulus is prevented, the process is asymptotically stable and the quantizer is not saturated. First, the preliminary result has shown that if the convergence rate of the pre-given closed-loop system is sufficiently fast, one can encrypt the controller without re-encryption. Subsequently, in the main result, the convergence rate of the pre-given closed-loop system is not required to be fast. 
 For a pre-given stabilizing controller, our scheme can ensure that there always exist a sufficiently small zooming-in factor of dynamic quantization that scales up all the controller's coefficients into integers, with which re-encryption is not required.

In the future, one may attempt to reduce the communication burdens by exploiting more historical data and the observability of the controller \cite{teranishi2023input,feng2024asymptotic}. It is also interesting to conduct experiments comparing the computation overheads in details, and reduce the computation overheads of the actuator by developing packing techniques\cite{jang2024ring}.  

\appendix
\textbf{Proof of Proposition \ref{proposition 1}.}
First note that the right-hand side of (\ref{15}a) is equivalent to $\tilde u(t) - \Big\lfloor \frac{\tilde u(t)  - \frac{\tilde u(t-1)}{\omega} + \frac{q}{2}}{q} \Big\rfloor  q$. Then we are interested in the following question: whether $\tilde u(t)  - \frac{\tilde u(t-1)}{\omega}= [JC\,\,\, H][\bar x_p ^T (t)- \bar x_p ^T (t-1)/\omega\,\,\,\,\bar x ^T  (t)- \bar x^T (t-1)/\omega]^T=[JC\,\,\, H] \delta(t)$ has an upper bound. 
The dynamics of $\delta(t)$ follows
\begin{align}\label{delta}
	&\delta(t+1)
	=\frac{1}{\omega}
	\underbrace{\begin{bmatrix}
			A+BJC & BH\\
			GC& F
	\end{bmatrix}}_{:=A_{\text{cl}}}
	\delta(t) \nonumber\\
	&\quad\quad + \frac{1}{\omega}
	\begin{bmatrix}
		BJ & B & BS\\ G & 0 & R
	\end{bmatrix}
	\underbrace{\begin{bmatrix}
			\bar e_{\bar y}(t)- \frac{\bar e_{\bar y}(t-1)}{\omega} \\ 
			\bar  e_{\bar u}(t)- \frac{\bar e_{\bar u}(t-1)}{\omega}\\
			\bar e_{\bar r}(t) -\frac{\bar e_{\bar r}(t-1)}{\omega}
	\end{bmatrix}}_{:=\bar e(t)}
\end{align}
in which the quantization errors are denoted by $\bar e_{\bar y}(t)= Q(\bar y (t))- \bar y(t)$, $\bar e_{\bar u}(t)= Q(\bar u (t))- \bar u(t)$ and $\bar e_{\bar r}(t)= Q(\bar r (t))- \bar r(t)$.

The following three aspects imply that $\delta(t)$ in (\ref{delta}) has an upper bound. 1) By $\delta(0)= [\bar x_p ^T (0)- \bar x_p ^T (-1)/\omega\,\,\,\,\bar x ^T  (0)- \bar x^T (-1)/\omega]^T $, one can inter that $\|\delta(0)\|$ has an upper bound, assuming that the initial conditions $x_p(0)$, $x_p(-1)$, $x(0)$ and $x(-1)$ are not infinitely large. 2) Because we have assumed that the quantizer has an infinite quantization range in this subsection, it is clear that $\|\bar e_{\bar y}(t)\|_\infty \le \frac{1}{2}$, $\|\bar e_{\bar u}(t)\|_\infty \le \frac{1}{2}$ and $\|\bar e_{\bar r}(t)\|_\infty \le \frac{1}{2}$. Then, one can verify $\|\bar e(t)\|\le \sqrt{v+w+n_r}(\frac{1}{2}+\frac{1}{2\omega})$. 3) In view of the value of $\omega$ in Lemma \ref{lemma}, one has $\rho(A_{\text{cl}}/\omega)<1$. 
By 1)--3), one can infer that there exists $M\in \mathbb{R}_{>0}$ such that $\|\delta(t)\|\le M$. 
Therefore, one has $\|\tilde u(t)  - \frac{\tilde u(t-1)}{\omega}\|_\infty \le \|[JC\,\,\, H]\|_\infty \|\delta(t)\| \le \|[JC\,\,\, H]\|_\infty M$.

We let $q> 2\|[JC\,\,\, H]\|_\infty M$. By such a $q$, it is clear that $ \Big\lfloor \frac{\tilde u(t)  - \frac{\tilde u(t-1)}{\omega} + \frac{q}{2}}{q} \Big\rfloor = \mathbf 0$ and hence $\tilde u_a(t) = \tilde u(t)$. Moreover, because $u(t)= s_1 s_2 l(t) \tilde u(t)$ by definition, eventually one can conclude that $u_a(t)= u(t)$, i.e., the actuator restores the control input in (\ref{5}). \qedp


\bibliographystyle{IEEEtran}

\bibliography{bib}

\begin{thebibliography}{10}
\providecommand{\url}[1]{#1}
\csname url@samestyle\endcsname
\providecommand{\newblock}{\relax}
\providecommand{\bibinfo}[2]{#2}
\providecommand{\BIBentrySTDinterwordspacing}{\spaceskip=0pt\relax}
\providecommand{\BIBentryALTinterwordstretchfactor}{4}
\providecommand{\BIBentryALTinterwordspacing}{\spaceskip=\fontdimen2\font plus
\BIBentryALTinterwordstretchfactor\fontdimen3\font minus
  \fontdimen4\font\relax}
\providecommand{\BIBforeignlanguage}[2]{{%
\expandafter\ifx\csname l@#1\endcsname\relax
\typeout{** WARNING: IEEEtran.bst: No hyphenation pattern has been}%
\typeout{** loaded for the language `#1'. Using the pattern for}%
\typeout{** the default language instead.}%
\else
\language=\csname l@#1\endcsname
\fi
#2}}
\providecommand{\BIBdecl}{\relax}
\BIBdecl

\bibitem{sisinni2018industrial}
E.~Sisinni, A.~Saifullah, S.~Han, U.~Jennehag, and M.~Gidlund, ``Industrial
  internet of things: Challenges, opportunities, and directions,'' \emph{IEEE
  Transactions on Industrial Informatics}, vol.~14, no.~11, pp. 4724--4734,
  2018.

\bibitem{schluter2023brief}
N.~Schl{\"u}ter, P.~Binfet, and M.~S. Darup, ``A brief survey on encrypted
  control: From the first to the second generation and beyond,'' \emph{Annual
  Reviews in Control}, p. 100913, 2023.

\bibitem{kim2022comparison}
J.~Kim, D.~Kim, Y.~Song, H.~Shim, H.~Sandberg, and K.~H. Johansson,
  ``Comparison of encrypted control approaches and tutorial on dynamic systems
  using learning with errors-based homomorphic encryption,'' \emph{Annual
  Reviews in Control}, vol.~54, pp. 200--218, 2022.

\bibitem{darup2021encrypted}
M.~S. Darup, A.~B. Alexandru, D.~E. Quevedo, and G.~J. Pappas, ``Encrypted
  control for networked systems: An illustrative introduction and current
  challenges,'' \emph{IEEE Control Systems Magazine}, vol.~41, no.~3, pp.
  58--78, 2021.

\bibitem{kogiso2015cyber}
K.~Kogiso and T.~Fujita, ``Cyber-security enhancement of networked control
  systems using homomorphic encryption,'' in \emph{IEEE Conference on Decision
  and Control}, 2015, pp. 6836--6843.

\bibitem{shi2024quantization}
Y.~Shi and E.~Nekouei, ``Quantization and event-triggered policy design for
  encrypted networked control,'' \emph{IEEE/CAA Journal of Automatica Sinica},
  vol.~11, no.~4, pp. 946--955, 2024.

\bibitem{darup2017towards}
M.~S. Darup, A.~Redder, I.~Shames, F.~Farokhi, and D.~Quevedo, ``Towards
  encrypted {MPC} for linear constrained systems,'' \emph{IEEE Control Systems
  Letters}, vol.~2, no.~2, pp. 195--200, 2017.

\bibitem{ruan2019secure}
M.~Ruan, H.~Gao, and Y.~Wang, ``Secure and privacy-preserving consensus,''
  \emph{IEEE Transactions on Automatic Control}, vol.~64, no.~10, pp.
  4035--4049, 2019.

\bibitem{alexandru2020cloud}
A.~B. Alexandru, K.~Gatsis, Y.~Shoukry, S.~A. Seshia, P.~Tabuada, and G.~J.
  Pappas, ``Cloud-based quadratic optimization with partially homomorphic
  encryption,'' \emph{IEEE Transactions on Automatic Control}, vol.~66, no.~5,
  pp. 2357--2364, 2020.

\bibitem{zhang2020secure}
Z.~Zhang, P.~Cheng, J.~Wu, and J.~Chen, ``Secure state estimation using hybrid
  homomorphic encryption scheme,'' \emph{IEEE Transactions on Control Systems
  Technology}, vol.~29, no.~4, pp. 1704--1720, 2020.

\bibitem{marcantoni2022secure}
M.~Marcantoni, B.~Jayawardhana, M.~P. Chaher, and K.~Bunte, ``Secure formation
  control via edge computing enabled by fully homomorphic encryption and mixed
  uniform-logarithmic quantization,'' \emph{IEEE Control Systems Letters},
  vol.~7, pp. 395--400, 2022.

\bibitem{kishida2019encrypted}
M.~Kishida, ``Encrypted control system with quantiser,'' \emph{IET Control
  Theory \& Applications}, vol.~13, no.~1, pp. 146--151, 2019.

\bibitem{farokhi2017secure}
F.~Farokhi, I.~Shames, and N.~Batterham, ``Secure and private control using
  semi-homomorphic encryption,'' \emph{Control Engineering Practice}, vol.~67,
  pp. 13--20, 2017.

\bibitem{cheon2018need}
J.~H. Cheon, K.~Han, H.~Kim, J.~Kim, and H.~Shim, ``Need for controllers having
  integer coefficients in homomorphically encrypted dynamic system,'' in
  \emph{IEEE Conference on Decision and Control}, 2018, pp. 5020--5025.

\bibitem{murguia2020secure}
C.~Murguia, F.~Farokhi, and I.~Shames, ``Secure and private implementation of
  dynamic controllers using semihomomorphic encryption,'' \emph{IEEE
  Transactions on Automatic Control}, vol.~65, no.~9, pp. 3950--3957, 2020.

\bibitem{schluter2021stability}
N.~Schl{\"u}ter and M.~S. Darup, ``On the stability of linear dynamic
  controllers with integer coefficients,'' \emph{IEEE Transactions on Automatic
  Control}, vol.~67, no.~10, pp. 5610--5613, 2021.

\bibitem{kimtac1}
J.~Kim, H.~Shim, and K.~Han, ``Dynamic controller that operates over
  homomorphically encrypted data for infinite time horizon,'' \emph{IEEE
  Transactions on Automatic Control}, vol.~68, no.~2, pp. 660--672, 2022.

\bibitem{teranishi2023input}
K.~Teranishi, T.~Sadamoto, and K.~Kogiso, ``Input--output history feedback
  controller for encrypted control with leveled fully homomorphic encryption,''
  \emph{IEEE Transactions on Control of Network Systems}, vol.~11, no.~1, pp.
  271--283, 2023.

\bibitem{tavazoei2022nonminimality}
M.~S. Tavazoei, ``Nonminimality of the realizations and possessing state
  matrices with integer elements in linear discrete-time controllers,''
  \emph{IEEE Transactions on Automatic Control}, vol.~68, no.~6, pp.
  3698--3703, 2022.

\bibitem{lee2023conversion}
J.~Lee, D.~Lee, S.~Lee, J.~Kim, and H.~Shim, ``Conversion of controllers to
  have integer state matrix for encrypted control: Non-minimal order
  approach,'' in \emph{IEEE Conference on Decision and Control}, 2023, pp.
  5091--5096.

\bibitem{kim2021method}
J.~Kim, H.~Shim, H.~Sandberg, and K.~H. Johansson, ``Method for running dynamic
  systems over encrypted data for infinite time horizon without bootstrapping
  and re-encryption,'' in \emph{IEEE Conference on Decision and Control}.\hskip
  1em plus 0.5em minus 0.4em\relax IEEE, 2021, pp. 5614--5619.

\bibitem{kimcdc}
J.~Kim, M.~S. Darup, H.~Sandberg, and K.~H. Johansson, ``Asymptotic
  stabilization over encrypted data with limited controller capacity and
  time-varying quantizer,'' in \emph{IEEE Conference on Decision and Control},
  2022, pp. 7762--7767.

\bibitem{feng2024asymptotic}
S.~Feng and J.~Kim, ``Asymptotic tracking control of dynamic reference over
  homomorphically encrypted data with finite modulus,'' \emph{arXiv preprint
  arXiv:2409.18787}, 2024.

\bibitem{Paillier}
P.~Paillier, ``Public-key cryptosystems based on composite degree residuosity
  classes,'' in \emph{Proceedings of the International Conference on Theory and
  Application of Cryptographic Techniques}, 1999, p. 223–238.

\bibitem{walsh2002stability}
G.~C. Walsh, H.~Ye, and L.~G. Bushnell, ``Stability analysis of networked
  control systems,'' \emph{IEEE transactions on Control Systems Technology},
  vol.~10, no.~3, pp. 438--446, 2002.

\bibitem{jang2024ring}
Y.~Jang, J.~Lee, S.~Min, H.~Kwak, J.~Kim, and Y.~Song, ``Ring-{LWE} based
  encrypted controller with unlimited number of recursive multiplications and
  effect of error growth,'' \emph{arXiv preprint arXiv:2406.14372}, 2024.

\end{thebibliography}

\end{document}